\documentstyle{article}
\begin{document}

\title{{\bf Alternative approach to the
regularization of odd dimensional AdS gravity }}
\author{Pablo Mora\footnote{E-mail: pablmora@gmail.com}
\\
{\it Instituto de F\'{\i}sica, Facultad de Ciencias}\\ {\it Igu\'a
4225, Montevideo, Uruguay}} \maketitle

\begin{abstract}
In this paper I present an action principle for odd dimensional
AdS gravity which consists of introducing another manifold with
the same boundary and a very specific boundary term. This new
action allows and alternative approach to the regularization of
the theory, yielding a finite euclidean action and finite
conserved charges.

The choice of the boundary term is justified on the grounds that an
enhanced 'almost off-shell' local AdS/Conformal symmetry arises for
that very special choice. One may say that the boundary term is
dictated by a guiding symmetry principle.

Two sets of boundary conditions are considered, which yield
regularization procedures analogous to (but different from) the
standard 'background substraction' and 'counterterms'
regularization methods.

The Noether charges are constructed in general. As an application it
is shown that for Schwarszchild-AdS black holes the charge
associated to the time-like Killing vector is finite and is indeed
the mass.\\
The Euclidean action for Schwarzschild-AdS black holes is computed,
and it turns out to be finite, and to yield the right
thermodynamics.

The previous paragraph may be interpreted in the sense that the
boundary term dictated by the symmetry principle is the one that
correctly regularizes the action.\\

PACS numbers: 0.450.+h, 11.10.Kk, 04.70.-s, 04.60.-m

\end{abstract}

\section{Introduction}

The AdS/CFT correspondence
\cite{maldacena1,maldacena2,witten-ads}\footnote{ A recent
overview of the status of the AdS-CFT correspondence is given in
Ref.\cite{horo}.} has originated a great interest in AdS gravities
for asymptotically AdS space-times. Semi-classical saddle point
calculations in the gravity side would correspond to strong
coupling properties of the dual conformal theory.

As the gravitational action diverges, suitable regularization
methods are required in order to obtain sensible results.

There are essentially two main approaches to the regularization
for the case of the standard Dirichlet boundary conditions, the
'background
substraction'\cite{gibbonshawking,hawkingpage,witten-ads} method
and the 'counterterms' method
\cite{skenderis,emparan,bala,deharo,skenderis1,papadimitriou}.

In this paper we present an alternative formalism for the
regularization of General Relativity with a cosmological constant
(also called 'AdS gravity') in odd dimensional space-times. To
that end I consider an action functional for odd dimensional AdS
gravity which consists of introducing another manifold with the
same boundary and a very specific boundary term. Two possible sets
of boundary conditions will be considered, different from the
standard Dirichlet boundary conditions, which lead to action
principles for which the action is regularized in a way analogous
to the 'counterterms' and to the 'background subtraction' methods
respectively, but differing from those methods in several aspects.

On the 'counterterms' side it is shown that the action and
formalism of ref.\cite{motz2} can be recovered as a particular
case. The approach of ref.\cite{motz2} has been studied and
developed further in refs.\cite{kounterterms1,kounterterms2},
where among other things the relationship with the above mentioned
standard counterterms approach is discussed. To avoid confusion
between the approach of
refs.\cite{motz2,kounterterms1,kounterterms2} and the counterterms
of
refs.\cite{skenderis,emparan,bala,deharo,skenderis1,papadimitriou}
R. Olea used the word 'kounterterms' to describe the former, a
convention that I will follow here.

The construction proposed in this paper also makes possible an
approach to the regularization in the spirit of the background
substraction methods, which however is not the same as the
Hawking-Page approach, as it will be shown below.

It is worthwhile to mention that the problem of regularizing AdS
gravity by introducing an action principle with suitable boundary
terms and non Dirichlet boundary conditions has been solved in the
even dimensional case in refs.\cite{aroscargas,aros}.

The idea of the alternative 'kounterterms' approach to the
regularization of odd-dimensional AdS gravity introduced in
ref.\cite{motz2,kounterterms1,kounterterms2} was to 'borrow' the
boundary term used to regularize Chern-Simons
gravities\footnote{For a review of Chern-Simons gravity see
ref.\cite{zanelli}.} in ref.\cite{motz1}, with a very specific
relative coefficient between the standard AdS gravity bulk term
and the 'borrowed' boundary term, and using the same boundary
conditions. Perhaps surprisingly such approach did work, yet one
can hardly avoid to wonder if there is some profound reason for
that to happen.

An important clue is that the boundary term in ref.\cite{motz1} can
be understood as coming from the extension from Chern-Simons forms
to transgression forms\footnote{For mathematical background on
transgression forms see refs.\cite{chern,eguchi,naka}.}, as
mentioned in ref.\cite{motz1} and discussed in detail in
refs.\cite{tesis,motz3}. Transgression forms involve two gauge
potentials $A$ and $\overline{A}$, and Chern-Simons forms are just
transgression forms with $\overline{A}=0$.

The extension of Chern-Simons to transgression forms as a device to
regularize the theory was done in 2+1 dimensions in
ref.\cite{postdam} (where the second field $\overline{A}$ was
understood as a fixed reference background), and in the context of
actions for extended objects in refs.\cite{mn,mora} (here with both
fields taken as dynamical). Afterwards, in the above mentioned
refs.\cite{motz1,tesis,motz3}, it was shown how to use
transgressions to properly regularize Chern-Simons gravity theories
in arbitrary dimension. Other works using transgressions as actions
are refs.\cite{BFF2003,BFFF2005,IRS,Aros-trans}.

The key point in the transition from Chern-Simons to transgression
forms is gauge invariance. While Chern-Simons forms are
quasi-invariant, changing by a closed form under gauge
transformations, transgression forms are truly gauge invariant. For
instance the motivation for that transition in refs.\cite{mn,mora}
was to have a truly gauge invariant action even in the case of
branes with boundaries. Thus the results of
refs.\cite{motz1,tesis,motz3} can be construed as follows: the
boundary terms dictated by the gauge principle turn out to be the
ones that properly regularize the action.

The question that naturally arises is then: is there a symmetry
principle which somehow explains why the boundary term of
ref.\cite{motz2} works? The answer is affirmative, and that is one
of the main results of this paper.

By analogy with the case of the transgressions for the AdS group
\cite{motz3} we introduce an action for odd-dimensional AdS gravity
with two sets of fields, that is that in addition to the vielbein
$e^a$ and the spin connection $\omega ^{ab}$ with support in a
manifold ${\cal M}$, we have a vielbein $\overline{e}^a$ and a spin
connection $\overline{\omega } ^{ab}$ with support in a manifold
$\overline{{\cal M}}$ with a common boundary with ${\cal M}$ (that
is $\partial{\cal M}\equiv\partial\overline{{\cal M}}$).

The complication of having an additional set of fields is
compensated by:

{\bf i.} The arising of an enhanced 'almost off-shell' local
AdS/Conformal symmetry for a very special choice of the boundary
term.

{\bf ii.} The fact that both the 'background subtraction' and the
'kounterterms' approach can be regarded as particular cases of
this framework. We show that the action principle of
ref.\cite{motz2} arises for a specific choice of the second field
(regarded as a 'reference configuration' or 'vacuum'). The
'background subtraction' regularization corresponds to a different
choice of the second field or 'reference configuration'. It is
however very important to emphasize again that in the
'kounterterms' case there is no extra input of information
required besides the original $A$ configuration, as the $\bar{A}$
is constructed from it in a direct way.

The Noether charges are constructed in general for the action
given, and the charge of ref.\cite{motz2}is a particular case of
the general formula.

As computations in the kounterterms side of the present framework
were done in ref.\cite{motz2,kounterterms1,kounterterms2}, here I
do some 'background subtraction' computations as an application,
and to show that the proposed method does indeed work. It is shown
that for Schwarszchild-AdS black holes the charge associated to
the time-like Killing vector is finite and is indeed the mass. The
Euclidean action for Schwarzschild-AdS black holes with different
horizon topologies is computed, with suitable backgrounds for each
case, and it turns out to be finite, and to yield the right
thermodynamics.

The important point of the relationship between the standard
counterterms approach and our kounterterms approach has been
discussed in refs.\cite{kounterterms1,kounterterms2} therefore I
refer the reader to those papers on that regard.

\section{\bf The action}

\subsection{General setting}

The reader be warned that the presentation of this section is
somehow indirect. I first write the AdS transgression form derived
in ref.\cite{motz3}, and the transgression action discussed there.
Afterwards I consider an action for AdS gravity with a boundary term
'borrowed' from the AdS transgression with a coefficient to be
determined, plus a doubling of he fields analogous to the one in the
transgression. We will show that with the proper coefficient the
resulting action has an enhanced symmetry and is properly
regularized.

\subsubsection{Review of AdS Transgressions}

We briefly review in this sub-subsection some results from
ref.\cite{motz3} that we will use in what follows.

The AdS transgression in dimension $d=2n+1$ is\footnote{I will use a
compact notation where $\epsilon$ stands for the Levi-Civita symbol
$\epsilon _{a_1...a_d}$ and wedge products of differential forms are
understood. For instance: $$\epsilon Re^{d-2}\equiv \epsilon
_{a_1a_2....a_d}R^{a_1a_2}\wedge e^{a_3}\wedge ...\wedge
e^{a_{d-2}}$$ This clarification should be enough to follow what
follows, but if there is any doubt on notation check
\cite{motz2,motz1,motz3}.} \cite{motz3}
\begin{equation}
{\cal T}_{2n+1}= \kappa \int _0^1dt \epsilon (R+t^2e^2)^ne -\kappa
\int _0^1dt \epsilon (\tilde{R}+t^2\overline{e}^2)^n\overline{e}
+d~\alpha _{2n}
\end{equation}
where
\begin{equation}
\alpha _{2n}=-\kappa n\int_0^1dt\int_0^1ds~\epsilon\theta e_t
\left\{t R+(1-t)\tilde{R}-t(1-t)\theta ^2+s^2e_t^2 \right\}^{n-1}
\end{equation}
Here $e ^a$ and $\overline{e}^a$ are the two vielbeins and $\omega
^{ab} $ and $\overline{\omega}^{ab}$ the two spin connections,
$R=d\omega +\omega ^2$ and
$\tilde{R}=d\overline{\omega}+\overline{\omega}^2$ are the
corresponding curvatures, $\theta =\omega -\overline{\omega}$ and
$e_t=te+(1-t)\overline{e}$. Written in a more compact way
\begin{equation}
\alpha _{2n}=-\kappa n\int_0^1dt\int_0^1ds~\epsilon\theta e_t
\overline{R} _{st}^{n-1}
\end{equation}
where
$$\overline{R} _{st}=t R+(1-t)\tilde{R}-t(1-t)\theta ^2+s^2e_t^2$$

The action for transgressions for the AdS group is taken to be
\cite{motz3}
\begin{equation}
I_{Trans}= \kappa \int _{\cal M}\int _0^1dt \epsilon (R+t^2e^2)^ne
-\kappa\int_{\overline{{\cal M}}} \int _0^1dt \epsilon
(\tilde{R}+t^2\overline{e}^2)^n\overline{e} +\int _{\partial {\cal
M} }\alpha _{2n}\label{transgresion}
\end{equation}
where ${\cal M}$ and $\overline{\cal M}$ are two manifolds with a
common boundary, that is the boundaries of ${\cal M}$ and
$\overline{{\cal M}}$ coincide $\partial {\cal M}\equiv\partial
\overline{{\cal M}}$. Notice that this is a generalization from he
simpler case where ${\cal M}\equiv\overline{\cal M}$, which is
physically motivated by he fact that both 'sheets' may no even have
the same topology (for instance if ${\cal M}$ is a black hole
spacetime and $\overline{\cal M}$ is the AdS spacetime).

We write the transgression action as
\begin{equation}
I_{Trans}= \int _{\cal M}L_{LLCS} -\int_{\overline{{\cal
M}}}\overline{L}_{LLCS} +\int _{\partial {\cal M} }\alpha _{2n}
\end{equation}
where the Lanczos-Lovelock-Chern-Simons lagrangian $L_{LLCS}$ is
\begin{equation}
L_{LLCS}= \kappa \int _0^1dt \epsilon (R+t^2e^2)^ne
\end{equation}

The transgression form is invariant under gauge transformations for
the AdS gauge group SO(d-2,2). These are the Lorentz transformations
\begin{equation}
\delta \omega ^{ab}=-d\lambda ^{ab}-\omega ^a_{~c}\lambda
^{cb}-\omega ^b_{~c}\lambda ^{ac}~~,~~\delta e^a=\lambda
^a_{~b}e^b
\end{equation}
and gauge translations
\begin{equation}
\delta \omega ^{ab}= e^b\lambda ^a-e^a\lambda ^b~~,~~\delta
e^a=-d\lambda ^a-\omega ^a_{~b}\lambda ^b
\end{equation}
Writing the AdS gauge connection as\footnote{Actually as a gauge
connection has dimensions of $(lenght)^{-1}$ we should write
$$A=\frac{\omega ^{ab}}{2}J_{ab}+\frac{e^a}{l}P_a$$ where $l$ is the 'AdS radius'.
We choose $l=1$ trough all the present paper, as it is
straightforward to reintroduce $l$ everywhere using dimensional
analysis.}
$$A=\frac{\omega ^{ab}}{2}J_{ab}+e^aP_a$$ where $J_{ab}$ and $P_a$
are the generators of the AdS group (for Lorentz transformations and
translations respectively) and the gauge parameter
$$\lambda=\frac{\lambda ^{ab}}{2}J_{ab}+\lambda ^aP_a$$ the AdS
gauge transformations take the compact form $$\delta A=-D\lambda
=-d\lambda -A\lambda +\lambda A $$ Invariance under Lorentz
transformations is immediate, as all the ingredients of the action
are Lorentz covariant and are contracted with the Levi-Civita
invariant tensor ($\theta =\omega -\overline{\omega}$ is Lorentz
covariant too, unlike $\omega$ or $\overline{\omega}$), but
invariance under translations is very non trivial, yet true.

The transgression action is also invariant under the AdS gauge
transformations, but it is only necessary that the transformations
on ${\cal M}$ and $\overline{\cal M}$ agree on $\partial{\cal M}$.
The variation of the boundary term cancels the variations of both
bulk terms.

\subsubsection{Extended action for General Relativity}

For General Relativity we consider the action
\begin{equation}
I_{GR}= \kappa \int _{\cal M}  \epsilon
\left[\frac{1}{d-2}Re^{d-2}+\frac{1}{d}e^d\right] -\kappa\int_{{\cal
\overline{M}}}\epsilon\left[\frac{1}{d-2}\tilde{R}\overline{e}^{d-2}+
\frac{1}{d}\overline{e}^d\right]+\chi _n\int _{\partial {\cal M}
}\alpha _{2n}\label{GR-action}
\end{equation}
with the same $\alpha _{2n}$. We will see in several ways that the
proper value for the constant $\chi _{n}$ is $\chi
_{n}=\frac{1}{n(d-2)f(n-1)}$ with $f(n-1)=\int
_0^1dt~(t^2-1)^{n-1}$. The action $I_{GR}$ is explicitly invariant
under Lorentz transformations, just like $I_{Trans}$. We write
$I_{GR}$ as
\begin{equation}
I_{GR}=\int _{\cal M}L_{EH} -\int_{{\cal \overline{M}}}
\overline{L}_{EH}+\chi _n\int _{\partial {\cal M} }\alpha _{2n}
\end{equation}
with the Einstein-Hilbert lagrangian $L_{EH}$ given by
\begin{equation}
L_{EH}= \kappa \epsilon
\left[\frac{1}{d-2}Re^{d-2}+\frac{1}{d}e^d\right]
\end{equation}
Notice he doubling of the fields analogous to the transgression. In
a saddle point evaluation of the euclidean action, or in the
evaluation of the Noether charges the configuration $A$ will be the
one of interest while the configuration $\overline{A}$ will be a
reference 'vacuum', giving the 'background substraction' or the
'counterterms' approach, depending on the nature of that 'vacuum'.

It is important however to emphasize that there is no 'little flag'
labeling one of the configurations as the dynamical one and the
other one as a background, and I believe in general both must be
treated in the same footing. In particular one could consider saddle
point configurations with the roles of $A$ and $\overline{A}$,
namely the 'configuration of interest' and 'the vacuum',
interchanged, considering the euclidean action with the opposite
sign\footnote{The idea is that he time variables of 'observers' in
each configuration must be regarded as having opposite signs, so
that each of them would see its 'piece' of the action as the
positive one, and none  would see undesirable effects as an
exponentially enhanced probability of black hole nucleation from its
past to its future (I am grateful to an anonymous referee for rising
this question).}.

\subsection{Variation of the action and field equations}

The variation of the transgression action yields
\begin{eqnarray}
\delta I_{Trans}= \kappa \int _{\cal M}\left[\epsilon
\overline{R}^n\delta
e+n\epsilon\overline{R}^{n-1}T\delta\omega\right] -\kappa \int
_{\cal \overline{M}}\left[\epsilon \overline{\tilde{R}}^n\delta
\overline{e}+n\epsilon\overline{\tilde{R}}^{n-1}\overline{T}\delta\overline{\omega}
\right]+\nonumber\\+ \int _{\partial {\cal M} } \left[
-n\kappa\int _0^1dt \epsilon (R+t^2e^2)^{n-1}e\delta\omega
+n\kappa \int _0^1dt \epsilon
(\tilde{R}+t^2\overline{e}^2)^{n-1}\overline{e}\delta\overline{\omega}
+\delta\alpha _{2n}\right]
\end{eqnarray}
where $\overline{R}=R+e^2$ and
$\overline{\tilde{R}}=\tilde{R}+\overline{e}^2$. From this variation
we can read the transgression field equations \cite{motz3}
\begin{eqnarray}
\epsilon
\overline{R}^n=0~~,~~\epsilon\overline{R}^{n-1}T=0\nonumber\\
\epsilon
\overline{\tilde{R}}^n=0~~,~~\epsilon\overline{\tilde{R}}^{n-1}\overline{T}=0
\end{eqnarray}
For general relativity the variation of the action is
\begin{eqnarray}
\delta I_{GR}= \kappa \int _{\cal M}\left[\epsilon
\overline{R}e^{d-3}\delta e+\epsilon Te^{d-3}\delta\omega\right]
-\kappa \int _{\cal \overline{M}}\left[\epsilon
\overline{\tilde{R}}\overline{e}^{d-3}\delta
\overline{e}+\epsilon\overline{T}\overline{e}^{d-3}\delta\overline{\omega}
\right]+\nonumber\\+ \int _{\partial {\cal M} } \left[ -\kappa
\epsilon  \frac{e^{d-2}}{d-2}\delta\omega +\kappa \epsilon
 \frac{\overline{e}^{d-2}}{d-2}\delta\overline{\omega}
+\delta\alpha _{2n}\right]\label{delta-GR-action}
\end{eqnarray}
giving the standard field equations for General Relativity with a
cosmological constant
\begin{eqnarray}
 \epsilon\overline{R}e^{d-3}=0~~,~~\epsilon Te^{d-3}=0\nonumber\\
\epsilon\overline{\tilde{R}}\overline{e}^{d-3}=0~~,~~
\epsilon\overline{T}\overline{e}^{d-3}=0
\end{eqnarray}
If the vielbein is invertible the equation of motion $\epsilon
Te^{d-3}=0$ implies $T=0$ (and
$\epsilon\overline{T}\overline{e}^{d-3}=0$ implies
$\overline{T}=0$).

It is of course clear that the equations of both theories are
quite different and that solutions for one theory will not in
general be solutions for the other.

\subsection{Boundary conditions}

When we obtained the field equations in the previous section, we
should have supplemented the action with suitable boundary
conditions that make the boundary contribution to the variation of
the action vanish, so that the action is truly an extremum when the
field equations hold. Here we present two such conditions, though
there may be others.

The first one, considered in refs.\cite{motz2,motz1}, which we
called {\it background independent configuration}, corresponds to
the reference configuration $\overline{A}$ chosen as
\begin{equation}
\overline{e}=0,~~~ \overline{\omega}^{ij}=\omega ^{ij}~~~and
~~~\overline{\omega}^{1j}=0,
\end{equation}
where $1$ corresponds to the direction normal to the boundary (the
normal being $e^1$) and $i,j$ are different from $1$. Then $\theta
^{ij}=0$ and $\theta ^{1i}=\omega ^{1i}$, $\overline{T}=0$ and
$\tilde{R}=R-\theta ^2$ for the components with support in the
boundary. This configuration is the most natural and economical
because given $e$ and $\omega $ no further information is required,
and the bulk term in $\overline{{\cal M}}$ vanishes, which can be
interpreted in the sense that the second manifold is not even
necessary. It is straightforward to check that for this
configuration the General Relativity action eq.(\ref{GR-action})
reduces to the one discussed in ref.\cite{motz2}, where it was shown
that it gives a well defined action principle for 'asymptotically
locally AdS' (ALAdS) space-times.

The second one, discussed for transgressions in ref.\cite{motz3},
corresponds to $$\Delta A\equiv A-\overline{A}\rightarrow 0$$ with a
fast enough fall-off to kill the boundary term when the coordinate
along the direction normal to the boundary approach the boundary.
Looking at eq.(\ref{delta-GR-action}) we see that in this case $\int
_{\partial {\cal M}}\alpha _{2n}\rightarrow 0$ just as in
ref.\cite{motz3}, while the remaining part of the boundary
contribution to the variation
$$\int _{\partial {\cal M} } \left[ -\kappa
\epsilon  \frac{e^{d-2}}{d-2}\delta\omega +\kappa \epsilon
 \frac{\overline{e}^{d-2}}{d-2}\delta\overline{\omega}\right]$$
will clearly vanish if $e-\overline{e}\rightarrow 0$ and $\omega
-\overline{\omega}\rightarrow 0$ fast enough towards the boundary,
as assumed. The fulfilment of this condition was explicitly checked
for the configurations considered in the concrete examples below.

I find it tempting to name the second condition 'boundary without
boundary' condition, as one may regard the manifolds ${\cal M}$ and
$\overline{{\cal M}}$ with opposite orientation (lets call it
$\overline{\cal M}^{*}$) joined at $\partial{\cal M}$ as a single
topological manifold ${\cal M}\bigcup \overline{\cal M}^{*}$, which
however would not be a smooth manifold in general. If the boundary
were at a finite distance boundary condition would mean that there
is no discontinuity across the boundary. In fact for a boundary at a
finite distance the fields would have no discontinuity and the
boundary term would be zero, meaning the boundary is a fictitious
one. I find the situation somewhat reminiscent of the 'method of
images' in electrostatic, where a physical situation with a boundary
is replaced by a configuration without a boundary in a wider region.

\subsection{Enhanced symmetry: local AdS symmetry for\\ General Relativity}

As mentioned before both $I_{Trans}$ and $I_{GR}$ are explicitly
invariant under Lorentz transformations. The transgression action
is in addition invariant off-shell under gauge translations and
hence under the whole AdS group. We will exploit that off-shell
invariance to show that the GR action with the boundary term given
and for a particular value of the coefficient $\chi _n$ is
invariant under the AdS group when certain conditions are
fulfilled.

We only need to consider gauge translations generated by the gauge
parameter $\lambda ^a$. The variation of $I_{Trans}$ is in this
case
\begin{eqnarray}
\delta _{\lambda} I_{Trans}= \int _{\partial {\cal M} } \{ -\kappa
\epsilon \overline{R}^n\lambda +\kappa \epsilon
\overline{\tilde{R}}^n\lambda -\nonumber\\+2n\kappa\int _0^1dt
\epsilon (R+t^2e^2)^{n-1}e^2\lambda -2n\kappa \int _0^1dt \epsilon
(\tilde{R}+t^2\overline{e}^2)^{n-1}\overline{e}^2\lambda +\delta
_{\lambda} \alpha _{2n}\}
\end{eqnarray}
But $\delta _{\lambda}I_{Trans}=0$ owing to the off-shell gauge
invariance of the transgression action, for any field
configuration. Then
\begin{eqnarray}
\int _{\partial {\cal M} }  \delta _{\lambda} \alpha _{2n}  = \int
_{\partial {\cal M} } \{ +\kappa \epsilon \overline{R}^n\lambda
-\kappa \epsilon \overline{\tilde{R}}^n\lambda
+\nonumber\\-2n\kappa\int _0^1dt \epsilon
(R+t^2e^2)^{n-1}e^2\lambda +2n\kappa \int _0^1dt \epsilon
(\tilde{R}+t^2\overline{e}^2)^{n-1}\overline{e}^2\lambda \}
\end{eqnarray}

On the other hand the variation of the GR action under gauge
translations is
\begin{eqnarray}
\delta _{\lambda} I_{GR}=\kappa \int_{{\cal M}}\epsilon
(d-3)\overline{R}Te^{d-4}\lambda -\kappa \int_{\overline{{\cal
M}}}\epsilon
(d-3)\overline{\tilde{R}}\overline{T}\overline{e}^{d-4}\lambda
+\nonumber\\+\int _{\partial {\cal M}}\{2\kappa\epsilon
\frac{e^{d-1}}{d-2}\lambda-2\kappa\epsilon
\frac{\overline{e}^{d-1}}{d-2}\lambda-\kappa\epsilon\overline{R}e^{d-3}\lambda
+\kappa\epsilon\overline{\tilde{R}}\overline{e}^{d-3}\lambda +\chi
_n\delta _{\lambda}\alpha _{2n} \}
\end{eqnarray}
where $\delta _{\lambda}\alpha _{2n}$ is given above.

We will ask for the following conditions:

{\bf (i)} Vanishing of the bulk terms
\begin{eqnarray}
 \epsilon (d-3)\overline{R}Te^{d-4}=0\\
\epsilon (d-3)\overline{\tilde{R}}\overline{T}\overline{e}^{d-4}=0
\end{eqnarray}
This conditions are certainly fulfilled if the torsions vanish
$T=\overline{T}=0$, which seems like a natural and rather weak
condition, but there may be other interesting configurations for
which the the torsion does not vanish but the bulk terms are still
zero, for instance some generalization to AdS gravity of the
configurations studied by Chand\'{\i}a and Zanelli \cite{chandia},
having $\overline{R}=\overline{\tilde{R}}=0$ but non zero torsions.

{\bf (ii)} Asymptotic solution to GR equations of motion. We
require that
\begin{eqnarray}
\int _{\partial {\cal M}}\{\epsilon\overline{R}e^{d-3}\lambda
-\epsilon\overline{\tilde{R}}\overline{e}^{d-3}\lambda \}=0
\end{eqnarray}
which is satisfied if the GR equations
$\epsilon\overline{R}e^{d-3}=0$ and
$\epsilon\overline{\tilde{R}}\overline{e}^{d-3}=0$ are satisfied
asymptotically (not necessarily in all the bulk) with a fast
enough fall off when approaching the boundary, but may also be
satisfied with weaker conditions owing to cancellation between
both terms.

{\bf (iii)} Asymptotically AdS configurations. We require that the
configuration are asymptotically AdS, $\overline{R}=0$ and
$\overline{\tilde{R}}=0$, or $R=-e^2$ and
$\tilde{R}=-\overline{e}^2$, with a fast enough fall off when
approaching the boundary to make $$\int _{\partial {\cal M} }\{
\epsilon \overline{R}^n\lambda - \epsilon
\overline{\tilde{R}}^n\lambda\}=0$$ and to make
\begin{eqnarray}
\int _{\partial {\cal M} }\{-2n\kappa\int _0^1dt \epsilon
(R+t^2e^2)^{n-1}e^2\lambda +2n\kappa \int _0^1dt \epsilon
(\tilde{R}+t^2\overline{e}^2)^{n-1}\overline{e}^2\lambda
\}=\nonumber\\ =\int _{\partial {\cal M} }\{-2n\kappa\int _0^1dt
\epsilon (t^2-1)^{n-1}e^{d-1}\lambda +2n\kappa \int _0^1dt
\epsilon (t^2-1)^{n-1}\overline{e}^{d-1}\lambda \}=\nonumber \\
=\int _{\partial {\cal M} }\{-2n\kappa f(n-1)\epsilon
e^{d-1}\lambda +2n\kappa f(n-1)\epsilon\overline{e}^{d-1}\lambda
\}
\end{eqnarray}
It is important to remark that condition (iii) does not imply
condition (ii), even though the AdS configurations are solutions
of the GR equations, because an asymptotic fall off fast enough to
make zero the terms required in (iii) may not be fast enough to
kill the terms required in (ii).

It turns out that if (i), (ii) and (iii) are satisfied and $\chi
_{n}=\frac{1}{n(d-2)f(n-1)}$ then the GR action is invariant under
gauge translations $$\delta _{\lambda}I_{GR}=0$$ which together
with its Lorentz invariance implies that under this conditions
$I_{GR}$ is invariant under the full AdS group (which is the
conformal group in $d-1$ dimensions).

Notice that the term $$\int_{\partial {\cal M}}\{2\kappa\epsilon
\frac{e^{d-1}}{d-2}\lambda-2\kappa\epsilon
\frac{\overline{e}^{d-1}}{d-2}\lambda \}$$ in $\delta _{\lambda}
I_{GR}$ is not automatically zero, so there is a non trivial
cancellation of this term with $\int _{\partial {\cal M}}\chi
_n\delta _{\lambda}\alpha _{2n}$. It is this very non trivial
cancellation what fixes the relative coefficient $\chi _n$ to the
unique value found, and there lies the heart of the invariance
presented in this subsection.\\

A slightly different case is the above mentioned {\it background
independent configuration}, considered in refs.\cite{motz2,motz1},
where $\overline{e}=0$, $\overline{\omega}^{ij}=\omega ^{ij}$ and
$\overline{\omega}^{1j}=0$, where $1$ corresponds to the direction
normal to the boundary (the normal being $e^1$) and $i,j$ are
different from $1$. Then $\theta ^{ij}=0$ and $\theta ^{1i}=\omega
^{1i}$, $\overline{T}=0$ and $\tilde{R}=R-\theta ^2$ for the
components with support in the boundary.

For the {\it background independent configuration} conditions (i)
and (ii) are the same, being even more easily fulfilled for
$\overline{e}$ and $\overline{\omega}$. The condition (iii) is
modified for $\overline{e}$ and $\overline{\omega}$ because
$\overline{\tilde{R}}=\tilde{R}$, and we require for those fields
$$\int _{\partial {\cal M} }\{ \tilde{R}^n\lambda\}=0$$
That is enough to make $$\delta _{\lambda}I_{GR}=0$$ which implies
that under this conditions $I_{GR}$ is invariant under the full AdS
group. It must be emphasized however that under generic AdS gauge
transformations the 'gauge condition' $\overline{e}=0$ is not
preserved.\\

The previous considerations are valid for the euclidean GR action
with periodic boundary conditions in the euclidean time, but also
for the lorentzian GR action provided that the gauge parameters in
the initial and final time space-like hypersurfaces vanish. The
vanishing of the gauge parameters for the initial and final
hypersurfaces in the lorentzian case is requiered because otherwise
the boundary terms of the variation coming from those hypersurfaces
would spoil the AdS invariance of the GR action.\\

In order to be more explicit about what it means a fast enough
fall off in conditions (ii) and (iii) we can be more specific
about the gauge parameter and afterwards look at the topological
black hole solutions considered in the next sections
\cite{topo1,topo2,scan} as examples to show that the class of
configurations satisfying both conditions is not only not empty
but rather quite wide.

We may consider a gauge parameter $\lambda$ which goes at most as
the radial coordinate $r$ as $r\rightarrow\infty$, that is
$\lambda ={\cal O}(r)$. For instance a parameter $\lambda$ such
that it is covariantly constant for AdS, $D\lambda =0$ does
satisfy that condition\footnote{In that case, with the coordinates
and notation of next section we should have $\lambda ^{1}=\lambda
^{0m}=0$, $\lambda ^{0}=\lambda ^{01}=C^{(1)}r$, $\lambda
^{m}=-\lambda ^{0m}=C^{(m)}r$ and $\lambda
^m_{~n}\tilde{e}^n=\omega ^m_{~n}C^{(m)}$, where the $C^{(a)}$'s
are arbitrary constants.}. In fact a dependence of $\lambda$ on a
higher power of $r$ will generate gauge transformations that are
singular at the boundary.

In that case condition (ii) is satisfied if
$\epsilon\overline{R}e^{d-3}$ and
$\epsilon\overline{\tilde{R}}\overline{e}^{d-3}$ fall off
asymptotically faster than $1/r$. The behaviour in the bulk is
clearly not constrained by condition (ii).

Concerning condition (iii), to make $$\int _{\partial {\cal M} }\{
\epsilon \overline{R}^n\lambda - \epsilon
\overline{\tilde{R}}^n\lambda\}=0$$  \footnote{Notice that in
these expressions $n=\frac{d-1}{2}$.}it will be necessary that
$\epsilon \overline{R}^n$ and $\epsilon \overline{\tilde{R}}^n$
would fall off faster than $1/r$. That is a quite weak condition
for asymptotically locally AdS space-times, easily met by the
black hole solutions considered in the next sections, for which
the components of $\overline{R}$ and $\overline{\tilde{R}}$ with
support at the boundary scale with $r$ as ${\cal O}(1/r^{d-3})$
(everything else in the integral being just 'angular factors').

Concerning the validity of eq.(23) what we need is that
asymptotically $\overline{R}=0$ and $\overline{\tilde{R}}=0$ with
a fast enough fall off to allow us to replace $R$ by $-e^2$ and
$\tilde{R}$ by $-\overline{e}^2$ in the integrals. The leading
order of the components with support at the boundary of $e^2$ and
$\tilde{e}^2$ go as $r^2$ for AdS asymptotics, and we have
$n-1=\frac{d-3}{2}$ factors $(R+t^2e^2)$ and a factor $e^2$ (or
$(\tilde{R}+t^2\tilde{e}^2)$ and a factor $\tilde{e}^2$). We can
write $R=\overline{R}-e^2$
($\tilde{R}=\overline{\tilde{R}}-\tilde{e}^2$), then if we suppose
$\lambda ={\cal O}(r)$ as before, the term with one $\overline{R}$
and $d-3$ $e$'s vanishes because of (ii) above (and the same holds
for the 'tilde' fields for all the paragraph) then the next order
corresponds to two $\overline{R}$ and $d-5$ $e$'s , to kill which
we must require that the components of $\overline{R}$ and
$\overline{\tilde{R}}$ with support at the boundary fall off with
$r$ faster than $1/r^{\frac{d-4}{2}}$.

As with condition (ii), condition (iii) does not impose any
restriction on the behaviour on the bulk.\\

If we consider as an example the black hole solutions of the next
sections, there is no problem with condition (ii), as those
configurations actually satisfy the field equations
$\epsilon\overline{R}e^{d-3}=0$ and
$\epsilon\overline{\tilde{R}}\overline{e}^{d-3}=0$, that is far
more than what we need to require.

The first part of condition (iii) is also verified, as it was just
mentioned. The second part of condition (iii) is also verified
because as we already said the components of $\overline{R}$ and
$\overline{\tilde{R}}$ with support at the boundary scale with $r$
as ${\cal O}(1/r^{d-3})$.\\

For the {\it background independent configuration} conditions (i)
and (ii) lead to the same requirements again. The condition (iii)
which is modified for $\overline{e}$ and $\overline{\omega}$ to
$$\int _{\partial {\cal M} }\epsilon\{ \tilde{R}^n\lambda\}=0$$
leads to a required fall off faster than $1/r$ for $\epsilon\{
\tilde{R}^n$ if $\lambda ={\cal O}(r)$, which is satisfied for the
above mentioned black hole configurations.

\section{Euclidean action and thermodynamics for Schwarszchild
black holes.}

In this section I will evaluate the euclidean action for
Schwarszchild black holes with different asymptotic topologies, with
suitable reference backgrounds. To that end I will first evaluate
the euclidean action with $A$ and $\overline{A}$ taken to be two
black hole like configurations with the same asymptotic topology,
and eventually chose the the right $\overline{A}$ so that for a
given black hole configuration $A$ the whole euclidean geometry is
non singular.

I will consider the action of eq.(\ref{GR-action}) and the black
hole solutions of refs.\cite{topo1,topo2,scan}. This solutions have
line element
\begin{equation}
ds^2=-\Delta ^2(r)dt ^2+\frac{dr^2}{\Delta ^2(r)}+r^2d\Sigma
_{d-2}^2
\end{equation}
with
\begin{equation}
\Delta ^2=\gamma-\frac{2GM}{r^{d-3}}+r^2
\end{equation}
where $d\Sigma _{d-2}^2$ is the line element of the
$(d-2)$-dimensional manifold of constant curvature proportional to
$\gamma =1,0,-1$. The event horizon $r_+$ is given by $\Delta
(r_+)=0$.

\subsection{Evaluation of the Euclidean Action}

In order to evaluate the euclidean action for two black hole
configurations with masses $M$ and $\overline{M}$ respectively the
relevant non vanishing ingredients are
\begin{eqnarray}
e^0=\Delta dt
~~,~~e^1=\frac{1}{\Delta}dr~~,~~e^m=r\tilde{e}^m\nonumber \\
\omega ^{01}=\left(\frac{\Delta ^2}{2}\right)'dt~~,~~ \omega
^{1m}=-\Delta \tilde{e}^m~~,~~\omega ^{mn}=\tilde{\omega
}^{mn}\nonumber \\ R^{01}=-\left(\frac{\Delta
^2}{2}\right)''dt~dr~~,~~R^{0m}=-\Delta \left(\frac{\Delta
^2}{2}\right)'dt\tilde{e}^m \nonumber \\ R^{1m}=-\frac{1}{\Delta
}\left(\frac{\Delta ^2}{2}\right)'dr\tilde{e}^m~~,~~R^{mn}=(\gamma
-\Delta ^2 )\tilde{e}^m\tilde{e}^n
\end{eqnarray}
with $$\Delta ^2=\gamma-\frac{2GM}{r^{d-3}}+r^2$$ (hence
$\left(\frac{\Delta ^2}{2}\right)'= r+(d-3)\frac{GM}{r^{d-2}}$),
for $\gamma =1,0,-1$ . Similar expressions hold with
$e\rightarrow\overline{e}$, $\omega\rightarrow\overline{\omega}$,
$R\rightarrow\tilde{R}$, $\Delta\rightarrow\overline{\Delta}$ and
$M\rightarrow\overline{M}$. We then have
\begin{eqnarray}
\theta ^{mn}=-(\Delta -\overline{\Delta })\tilde{e}^m~~,~~(\theta
^2 )^{mn}=-(\Delta -\overline{\Delta
})\tilde{e}^m\tilde{e}^n\nonumber\\ \theta
^{01}=\left[\left(\frac{\Delta
^2}{2}\right)'-\left(\frac{\overline{\Delta}
^2}{2}\right)'\right]dt ~~,~~(\theta ^2 )^{om}=-(\Delta
-\overline{\Delta })\left[\left(\frac{\Delta
^2}{2}\right)'-\left(\frac{\overline{\Delta}
^2}{2}\right)'\right]dt\tilde{e}^m\nonumber\\ e^0_t=[t\Delta
+(1-t)\overline{\Delta }]dt~~,~~e^m_t=r\tilde{e}^m\nonumber\\
(e^2_t)^{0m}=r(t\Delta +(1-t)\overline{\Delta })dt\tilde{e}^m~~,~~
(e^2_t)^{mn}=r^2\tilde{e}^m\tilde{e}^n
\end{eqnarray}
We will need the components of $\overline{R} _{st}=t
R+(1-t)\tilde{R}-t(1-t)\theta ^2+s^2e_t^2$ with group indices $mn$
and $0m$. Those are
\begin{eqnarray}
\left(  \overline{R} _{st}  \right)^{mn}=\left\{\gamma -[t\Delta
+(1-t)\overline{\Delta }]^2+s^2
r^2\right\}\tilde{e}^m\tilde{e}^n\nonumber\\ \left(\overline{R}
_{st}\right)^{0m}=\left\{ -t\Delta \left(\frac{\Delta
^2}{2}\right)'-(1-t)\overline{\Delta
}\left(\frac{\overline{\Delta} ^2}{2}\right)' +r(t\Delta
+(1-t)\overline{\Delta })s^2 \right\} dt\tilde{e}^m
\end{eqnarray}
The bulk contribution to the euclidean action can be evaluated
using the equations of motion $R+e^2=0$ and it is
\begin{equation}
I_E^{bulk}=2\kappa (d-3)!\beta \Sigma _{d-2} [r_+^{2n}]-2\kappa
(d-3)!\beta \Sigma _{d-2} [\overline{r}_+^{2n}]
\end{equation}
where $\Sigma _{d-2}$ is the volume of the constant curvature
manifold corresponding to the sections of fixed unity radius and
fixed euclidean time, which in the spherically symmetric case we
will also call $\Omega _{d-2}$, coming from integration over the
angular variables.

The boundary term is
\begin{eqnarray}
\alpha _{2n}=-\kappa n\int _0^1dt\int _0^1ds\{2\epsilon
_{1m_10m_2...m_{2n-1}}\theta ^{1m_1}e^0_t\overline{R}
_{st}^{m_2m_3}...\overline{R} _{st}^{m_{2n-2}m_{2n-1}}+\nonumber\\
+4(n-1)\epsilon _{1m_1m_20m_3...m_{2n-1}}\theta
^{1m_1}e^{m_2}_t\overline{R} _{st}^{0m_3}\overline{R}
_{st}^{m_4m_5}...\overline{R} _{st}^{m_{2n-2}m_{2n-1}}+
\nonumber\\ +2\epsilon _{01m_1m_2...m_{2n-1}}\theta
^{01}e^{m_1}_t\overline{R} _{st}^{m_2m_3}...\overline{R}
_{st}^{m_{2n-2}m_{2n-1}} \}
\end{eqnarray}
Inserting the expressions for the terms of this equation and taking
in account signs coming from bringing the $\epsilon$ to its standard
order, commuting differentials and an additional sign coming from
the orientation of the boundary (or equivalently bringing the
differential dr to the front from the canonical order
$dtdr\tilde{e}^{m_1}...\tilde{e}^{m _{2n-1}}\rightarrow
-drdt\tilde{e}^{m_1}...\tilde{e}^{m _{2n-1}}$ ) we get
\begin{eqnarray}
\int _{\partial {\cal M}}\alpha _{2n}= \kappa n2\beta (d-2)!\Sigma
_{d-2}\int _0^1dt\int _0^1ds\{ (\Delta -\overline{\Delta
})[t\Delta +(1-t)\overline{\Delta }]\times\nonumber\\
\times[\gamma -(t\Delta +(1-t)\overline{\Delta
})^2+s^2r^2]^{n-1}+\nonumber\\ 2(n-1)r(\Delta -\overline{\Delta
})[ -t\Delta \left(\frac{\Delta
^2}{2}\right)'-(1-t)\overline{\Delta
}\left(\frac{\overline{\Delta} ^2}{2}\right)' +r(t\Delta
+(1-t)\overline{\Delta })s^2]\times \nonumber \\\times [\gamma
-(t\Delta +(1-t)\overline{\Delta })^2+s^2r^2]^{n-2} +\nonumber\\
+\left[\left(\frac{\Delta
^2}{2}\right)'-\left(\frac{\overline{\Delta} ^2}{2}\right)'
\right]r[\gamma -(t\Delta +(1-t)\overline{\Delta
})^2+s^2r^2]^{n-1} \}
\end{eqnarray}
We will drop terms that give a vanishing contribution when
$r\rightarrow \infty$ and keep only divergent or finite
contributions in that limit. To that end we notice that
$$\left(\frac{\Delta ^2}{2}\right)'= r+(d-3)\frac{GM}{r^{d-2}}$$
and that for $r\rightarrow \infty$ we get $$\Delta\rightarrow
r+\frac{\gamma}{2r}-\frac{GM}{r^{d-2}}$$  and hence
\begin{eqnarray}
\left(\frac{\Delta ^2}{2}\right)'-\left(\frac{\overline{\Delta }
^2}{2}\right)'= (d-3)\frac{G(M-\overline{M})}{r^{d-2}}~~,~~ \Delta
-\overline{\Delta }\rightarrow
 -\frac{G(M-\overline{M})}{r^{d-2}}\nonumber\\
\gamma -(t\Delta +(1-t) \overline{\Delta })^2+s^2r^2 \rightarrow
(s^2-1)r^2+{\cal O}(r)\nonumber \\ (\Delta -\overline{\Delta
})\left[\left(\frac{\Delta
^2}{2}\right)'-\left(\frac{\overline{\Delta }
^2}{2}\right)'\right]\rightarrow {\cal O}(r^{-3(d-2)})\nonumber \\
r(\Delta -\overline{\Delta })\Delta \left(\frac{\Delta
^2}{2}\right)'\rightarrow r^2(\Delta -\overline{\Delta })\Delta
+{\cal O}(r^{-3(d-2)})\nonumber \\ r(\Delta -\overline{\Delta
})\overline{\Delta } \left(\frac{\overline{\Delta }
^2}{2}\right)'\rightarrow r^2(\Delta -\overline{\Delta
})\overline{\Delta } +{\cal O}(r^{-3(d-2)})
\end{eqnarray}
The boundary term is then
\begin{eqnarray}
\int _{\partial {\cal M}}\alpha _{2n}= \kappa n2\beta (d-2)!\Sigma
_{d-2}\int _0^1ds\{\int _0^1dt (\Delta -\overline{\Delta
})[t\Delta +(1-t)\overline{\Delta }]\times\nonumber\\
\times[\gamma -(t\Delta +(1-t)\overline{\Delta
})^2+s^2r^2]^{n-1}+\nonumber\\ 2(n-1)r^2(s^2-1)(\Delta
-\overline{\Delta })(t\Delta +(1-t)\overline{\Delta })\times
\nonumber \\\times [\gamma -(t\Delta +(1-t)\overline{\Delta
})^2+s^2r^2]^{n-2} \}+\nonumber\\ +\kappa n2\beta (d-2)!\Sigma
_{d-2}\int _0^1ds~(d-3)G(M-\overline{M})(s^2-1)^{n-1}
\end{eqnarray}
The integral in the parameter $t$ can be done trough the
substitution $$u=\gamma -(t\Delta +(1-t)\overline{\Delta
})^2+s^2r^2$$ and the result is
\begin{eqnarray}
\int _{\partial {\cal M}}\alpha _{2n}=-\kappa \beta (d-2)!\Sigma
_{d-2} \int _0^1ds[u^n+2nr^2(s^2-1)u^{n-1}]\mid _{\gamma
-\overline{\Delta } ^2+s^2r^2}^{\gamma -\Delta ^2+s^2r^2}
+\nonumber\\ +\kappa n2\beta (d-2)!\Sigma _{d-2}\int
_0^1ds~(d-3)G(M-\overline{M})(s^2-1)^{n-1}
\end{eqnarray}
Notice that $\gamma -\Delta
^2+s^2r^2=(s^2-1)r^2+\frac{2GM}{r^{d-3}}$ and $1-\overline{\Delta
} ^2+s^2r^2=(s^2-1)r^2+\frac{2G\overline{M}}{r^{d-3}}$ and that
$\alpha _{2n}$ is evaluated at the boundary where
$r\rightarrow\infty$. Keeping only terms that will give a
divergent or finite contribution we can expand
\begin{eqnarray}
\left[(s^2-1)r^2+\frac{2GM}{r^{d-3}}\right]^{n}=(s^2-1)^nr^{2n}
+n2GM(s^2-1)^{n-1}+...\nonumber\\
\left[(s^2-1)r^2+\frac{2GM}{r^{d-3}}\right]^{n-1}=(s^2-1)^{n-1}r^{2n-2}
+(n-1)\frac{2GM}{r^2}(s^2-1)^{n-2}+...
\end{eqnarray}
The divergent contributions cancel between the upper and lower
limits of the integrals. The resulting $\alpha _{2n}$ is then
\begin{equation}
\int _{\partial {\cal M}}\alpha _{2n}=-\kappa \beta (d-2)!\Sigma
_{d-2}f(n-1)n2G(M-\overline{M})
\end{equation}
where $$f(n-1)=\int _0^1ds (s^2-1)^{n-1}=(-1)^{n-1}
~\frac{(n-1)!2^{n-1}}{(2n-1)!!}$$ With the choice $\chi
_{n}=\frac{1}{n f(n-1)(d-2)}$ the total action reads
\begin{equation}
I_E^{Total}=2\kappa (d-3)!\beta \Sigma _{d-2}\{ [r_+^{2n}]-
[\overline{r}_+^{2n}]-G(M-\overline{M})\}
\end{equation}
We can replace $\kappa =\frac{1}{2G(d-2)!\Omega _{d-2}}$ to get
\begin{equation}
I_E^{Total}=\beta\frac{\Sigma _{d-2} }{\Omega _{d-2}
}\frac{1}{(d-2)G} \{ [r_+^{d-1}]-
[\overline{r}_+^{d-1}]\}-\beta\frac{\Sigma _{d-2} }{\Omega _{d-2}
}\frac{(M-\overline{M})}{(d-2)}\
\end{equation}

\subsection{Determination of $\beta$ and a suitable reference background}

The euclidean time period $\beta $ is determined by requiring that
the euclidean solution be non singular. For generic black hole
metrics of the form $$ds^2=\Delta ^2(r)dt ^2+\frac{dr^2}{\Delta
^2(r)}+r^2d\Sigma _{d-2}^2$$ with the an event horizon $r_+$ given
by $\Delta (r_+)=0$ it turns out that $$\beta =\frac{4\pi}{(\Delta
^2)'(r_+)}$$ For a General Relativity  black hole of mass $M$ it is
$$\beta =\frac{4\pi}{(d-1)r_++\frac{(d-3)\gamma }{r_+}}$$ This
implies $$M=\frac{r_+^{d-3}}{2G}(\gamma +r_+^2)$$ In order to have a
sensible thermodynamics the background configuration must be chosen
as having an arbitrary $\overline{\beta}$. For $\gamma=0$ and
$\gamma=1$ the proper configurations are the zero mass black holes,
with $\overline{M}=0$ and $\overline{r}_+=0$, which correspond to
AdS for $\gamma =1$. This configurations have
$\overline{\beta}=\infty$, what amounts to an ill-defined or
arbitrary $\overline{\beta}$, as it can be checked that this
euclidean configurations are non singular for any
$\overline{\beta}$. In particular, for the evaluation of the
euclidean action of the previous section to make sense we must take
$\overline{\beta}=\beta$.

In the case $\gamma =-1$ the situation is a little more
complicated. Requiring an ill-defined $\overline{\beta}=\infty$
yields in this case $$\overline{r}_+=\pm \sqrt{\frac{d-3}{d-1}}$$
and as $r$ is positive we must pick the positive root. This value
of $\overline{r}_+$ gives $$\overline{M}=-\frac{1}{(d-1)G}
\left[\frac{d-3}{d-1}\right]^{\frac{d-3}{2}}\equiv M_0$$ Notice
that $M_0$ is negative. Again for the evaluation of the euclidean
action to make sense we need to take  $\overline{\beta}=\beta$.

\subsection{Black hole thermodynamics}

We then get
\begin{equation}
I_E^{Total}=\beta\frac{\Sigma _{d-2} }{\Omega _{d-2}
}\frac{1}{(d-2)G} [r_+^{d-1}] -\beta\frac{\Sigma _{d-2} }{\Omega
_{d-2} }\frac{M }{(d-2)}+\beta\frac{\Sigma _{d-2} }{\Omega _{d-2}
}M_0\delta_{-1,\gamma}
\end{equation}
which coincides with the result obtained in \cite{motz2}, except for
a different $M$ independent term (the one that yields the vacuum
energy) in that case.

To discuss the black hole thermodynamics we use that the euclidean
action $I$ is related with the free energy $F$ as $I=-\beta F$,
while the free energy is related to the energy $E$ and the entropy
$S$ as $F=E-TS=E-S/\beta $. Equivalently
\begin{equation}
I=-\beta E+S
\end{equation}
Hence
\begin{equation}
E=-\frac{\partial I}{\partial \beta }=-\frac{\frac{\partial
I}{\partial r_{+}}}{\frac{\partial\beta}{\partial r_{+}}}
\end{equation}
The result of this calculation is
\begin{equation}
E=\frac{\Sigma _{d-2} }{\Omega _{d-2} }(M-M_0\delta_{-1,\gamma})
\end{equation}
The entropy can be calculated from
\begin{equation}
S=I+\beta E
\end{equation}
and it is
\begin{equation}
S=\frac{2\pi r_{+}^{2n-1}}{(2n-1)G}\frac{\Sigma _{d-2}}{\Omega
_{d-2}} =\frac{r_{+}^{2n-1}\Sigma _{d-2}}{4G_N}=\frac{A}{4G_N}
\end{equation}
in agreement with the Bekenstein-Hawking entropy. We used that the
standard Newton constant $G_N$ and the constant $G$ are related as
\cite{scan}
$$G=\frac{8\pi }{(d-2)\Omega _{d-2}}G_N$$

\subsection{Discussion: boundary terms versus the Hawking-Page approach}

As it was already pointed out in ref.\cite{motz3}  the background
subtraction procedure used here is not the same as the one
proposed by Gibbons and Hawking \cite{gibbonshawking}, or by
Hawking and Page \cite{hawkingpage}. In those papers, the actions
for two different configurations (for instance, for a black hole
and Minkowski or AdS space) are subtracted, with the additional
condition that the metrics match at a very large finite radius
$r_0$ (eventually taken to infinity). In that case two different
euclidean time intervals $\beta$ and $\overline{\beta}$ are
involved, because the condition $$ds^2\mid
_{r_0}=\overline{ds}^2\mid _{r_0}$$ implies $$\Delta (r_0)\beta
(r_0) = \overline{\Delta}(r_0)\overline{\beta} (r_0) $$ then, even
though $\overline{\beta}\rightarrow\beta $ when
$r_0\rightarrow\infty$, there is an extra contribution to the
total bulk action (the difference of the bulk actions for the
configuration of interest and the 'background') coming from the
difference of the $\beta$'s \cite{hawkingpage,witten-ads}.

In our approach there is always only one $\beta$, as it must be in
order to integrate the boundary term $B_{2n}$, where both sets of
vielbein and spin connections appear entangled, but we do have an
extra contribution coming from that boundary term.

It is worthwhile to emphasize that boundary term contributions are
absent in the Hawking-Page approach to asymptotically AdS
space-times, as the Gibbons-Hawking term is zero in that case.

It is instructive to compare both methods for the concrete example
of the Schwarzschild-AdS black hole with spherical symmetry, as the
extra contribution coming in the Hawking-Page method from the
differing $\beta$'s comes in our approach from the boundary term,
and both methods agree in that case.

When it comes to the conserved charges, discussed in the next
section, we will see for the concrete example of the Schwarszchild
black hole that the boundary term contribution is necessary to
obtain a result for the mass in agreement with the one obtained from
the thermodynamics. See in particular Section 4.2 below and the
comment in the last paragraph of that section. It is hard to see
where could this contribution come from in the Hawking-Page
approach, as the Noether charges are defined as integrals in the
boundary of spatial sections, therefore no integral in the euclidean
time is done and $\beta$ or $\overline{\beta}$ are not involved in
the result.

\section{Conserved charges from Noether's theorem}

\subsection{Noether's charges}

The action is
\begin{equation}
I= \kappa \int _{\cal M}  \epsilon
\left[\frac{1}{d-2}Re^{d-2}+\frac{1}{d}e^d\right]
-\kappa\int_{{\cal
\overline{M}}}\left[\frac{1}{d-2}\tilde{R}\overline{e}^{d-2}+
\frac{1}{d}\overline{e}^d\right]+\chi _n\int _{\partial {\cal M}
}\alpha _{2n}
\end{equation}
where
\begin{equation}
\alpha _{2n}=-\kappa n\int_0^1dt\int_0^1ds~\epsilon\theta e_t
\left\{t R+(1-t)\tilde{R}-t(1-t)\theta ^2+s^2e_t^2 \right\}^{n-1}
\end{equation}
$\chi _n$ is a constant relative factor and the boundaries
coincide $\partial {\cal M}\equiv\partial {\cal \overline{M}}$.
Applying Noether's theorem to this action we get the conserved
current associated to the invariance under diffeomorphisms
generated by the vector field $\xi$
\begin{equation}
\star j=dQ_{\xi}
\end{equation}
 with\footnote{The contraction operator $I_{\xi }$ is defined by
 acting on a p-form $\alpha _p$ as
$$
I_{\xi }\alpha _{p}=\frac{1}{(p-1)!}\xi ^{\nu }\alpha _{\nu \mu
_{1}...\mu _{p-1}}dx^{\mu _{1}}...dx^{\mu _{p-1}}
$$
and being and anti-derivative in the sense that acting on the wedge
product of differential forms $\alpha _{p}$ and $\beta _{q}$ of
order p and q respectively gives $I_{\xi }(\alpha _{p}\beta
_{q})=I_{\xi }\alpha _{p}\beta _{q}+(-1)^{p}\alpha _{p}I_{\xi }\beta
_{q}$.}
\begin{equation}
Q_{\xi}=\frac{\kappa}{(d-2)}\epsilon [e^{d-2}I_{\xi}\omega
-\overline{e}^{d-2}I_{\xi}\overline{\omega } ]+\chi
_nI_{\xi}\alpha _{2n}
\end{equation}
which is to be integrated at the spatial boundary $\partial {\cal
S}$, which for instance for topological black holes is $\Sigma
^{d-2}$. Here
\begin{eqnarray}
I_{\xi}\alpha _{2n}=-\kappa
n\epsilon\{\int_0^1dt\int_0^1ds~I_{\xi}\theta e_t
\overline{R}_{st}^{n-1} -\int_0^1dt\int_0^1ds~\theta I_{\xi}e_t
\overline{R}_{st}^{n-1}+\nonumber\\
+(n-1)\int_0^1dt\int_0^1ds~\theta e_t
I_{\xi}\overline{R}_{st}\overline{R}_{st}^{n-2} \}
\end{eqnarray}
where $\overline{R} _{st}=t R+(1-t)\tilde{R}-t(1-t)\theta
^2+s^2e_t^2$.

\subsection{Black hole mass}

We will evaluate the charge corresponding to $\xi =\frac{\partial
~ }{\partial t}$ for two black hole configurations with masses $M$
and $\overline{M}$ respectively the relevant non vanishing
ingredients are the same used in the evaluation of the euclidean
action for black holes. We have
\begin{equation}
\int _{\partial {\cal S}}\frac{\kappa}{(d-2)}\epsilon
[e^{d-2}I_{\xi}\omega -\overline{e}^{d-2}I_{\xi}\overline{\omega }
]=\frac{\kappa}{(d-2)}\Sigma_{d-2}(d-2)!r^{d-2}\left[\left(\frac{\Delta
^2}{2}\right)'-\left(\frac{\overline{\Delta} ^2}{2}\right)'\right]
\end{equation}
At the boundary $r\rightarrow\infty$ we get
\begin{equation}
\int _{\partial {\cal S}}\frac{\kappa}{(d-2)}\epsilon
[e^{d-2}I_{\xi}\omega -\overline{e}^{d-2}I_{\xi}\overline{\omega }
]=\frac{(d-3)}{(d-2)}\kappa\Sigma_{d-2}(d-2)! 2G(M-\overline{M})
\end{equation}
Furthermore we have the following non vanishing contribution to
$I_{\xi}\alpha _{2n}$
\begin{eqnarray}
I_{\xi}\alpha _{2n}=-\kappa n\int _0^1dt\int _0^1ds\{2\epsilon
_{01m_1...m_{2n-1}}I_{\xi}\theta ^{01}e^{m_1}_t\overline{R}
_{st}^{m_2m_3}...\overline{R} _{st}^{m_{2n-2}m_{2n-1}}-\nonumber\\
-2\epsilon _{1m_10m_2m_3...m_{2n-1}}\theta ^{1m_1}I_{\xi}e^{0}_t
\overline{R} _{st}^{m_2m_3}...\overline{R}
_{st}^{m_{2n-2}m_{2n-1}}+ \nonumber\\ +4(n-1)\epsilon
_{1m_1m_20m_3...m_{2n-1}}\theta
^{1m_1}e^{m_2}_tI_{\xi}\overline{R} _{st}^{0m_3}\overline{R}
_{st}^{m_4m_5}...\overline{R} _{st}^{m_{2n-2}m_{2n-1}} \}
\end{eqnarray}
Inserting the expressions for the terms of this equation and
taking in account signs coming from bringing the $\epsilon$ to its
standard order we get
\begin{eqnarray}
\int _{\partial {\cal S}}I_{\xi}\alpha _{2n}= -2\kappa n
(d-2)!\Sigma _{d-2}\int _0^1dt\int
_0^1ds\{\left[\left(\frac{\Delta
^2}{2}\right)'-\left(\frac{\overline{\Delta} ^2}{2}\right)'
\right]\times\nonumber\\ \times r[\gamma -(t\Delta
+(1-t)\overline{\Delta })^2+s^2r^2]^{n-1} +\nonumber\\ + (\Delta
-\overline{\Delta })[t\Delta +(1-t)\overline{\Delta }][\gamma
-(t\Delta +(1-t)\overline{\Delta })^2+s^2r^2]^{n-1} +\nonumber\\
+2(n-1)r(\Delta -\overline{\Delta })[ -t\Delta \left(\frac{\Delta
^2}{2}\right)'-(1-t)\overline{\Delta
}\left(\frac{\overline{\Delta} ^2}{2}\right)' +r(t\Delta
+(1-t)\overline{\Delta })s^2]\times \nonumber \\\times [\gamma
-(t\Delta +(1-t)\overline{\Delta })^2+s^2r^2]^{n-2} \}
\end{eqnarray}
Notice that the integrals in $s$ and $t$ are just the same we did
in the evaluation of the euclidean action, then we can directly
write down the result
\begin{equation}
\int _{\partial {\cal S}}I_{\xi}\alpha _{2n}=\kappa (d-2)!\Sigma
_{d-2}f(n-1)n2G(M-\overline{M})
\end{equation}
where $f(n-1)=\int _0^1ds
(s^2-1)^{n-1}=(-1)^{n-1}\frac{(n-1)!2^{n-1}}{(2n-1)!!}$. With the
choice $\chi _{n}=\frac{1}{n f(n-1)(d-2)}$ the total charge reads
\begin{equation}
\int _{\partial {\cal S}}Q_{\xi}=\kappa\Sigma_{d-2}(d-2)!
2G(M-\overline{M})
\end{equation}
We can replace $\kappa =\frac{1}{2G(d-2)!\Omega _{d-2}}$ to get
\begin{equation}
\int _{\partial {\cal S}}Q_{\xi}= \frac{\Sigma_{d-2}}{\Omega
_{d-2}} (M-\overline{M})
\end{equation}
which is the expected result. Notice that without the
$I_{\xi}\alpha _{2n}$ contribution the charge would be
\begin{equation}
\int _{\partial {\cal S}}Q^{bulk}_{\xi}=\frac{(d-3)}{(d-2)}
\frac{\Sigma_{d-2}}{\Omega _{d-2}} (M-\overline{M})
\end{equation}
It is worthwhile to emphasize the significance of this result: the
contribution to the Noether's  conserved charge coming from the
boundary term is {\bf required} to get a value of the mass in
agreement with the one coming from he thermodynamics.

\section{Discussion and Conclusions}

The results of this paper show that the boundary term suitable for
properly regularizing odd-dimensional AdS gravity may be regarded as
dictated or suggested by a symmetry principle, which could be seen
as the reason for that particular term to work. The symmetry
principle invoked is invariance under local transformations of the
AdS group, and it holds 'almost off-shell'.

The calculations of the thermodynamics and Noether charges of
AdS-Schwarszchild black holes with different topologies are new
not in the results, but are quite different in the methods used,
as the contribution of the boundary terms is crucial in our
approach, as discussed in section 3.4 and the last paragraph of
section 4.2. Our method provides a regularization procedures which
allows those calculations to be made in a uniform and
 systematic way for any solution.

The present work rises several questions and could be extended in
several directions:

The fact that the action considered, with the precise boundary
term chosen, has that extra symmetry suggest that it may be
relevant in the study of the AdS-CFT correspondence, as the AdS
group in dimension $d$ is the conformal group in dimension $d-1$
\footnote{If that were the case, it would however be puzzling that
the conformal symmetry that would be induced if one chooses
boundary conditions at infinity that do not break the symmetry and
integrates out the bulk degrees of freedom would have a local
symmetry with the conformal group as gauge group, while the CFT
side of the AdS-CFT correspondence involves a globally symmetric
conformal field theory. It may be that integrating out the bulk
degrees of freedom corresponding to $A$ while keeping the degrees
of freedom associated to the configuration $\overline{A}$ of
ref.\cite{motz2,motz1} as boundary degrees of freedom of the
effective theory would reduce the symmetry from a local gauge
redundancy to a global symmetry}. It is clear however that the
possible relevance of the setup considered here to the AdS-CFT
correspondence, beyond the remarks of ref.\cite{motz2}, is just an
interesting open question, which would require to address several
issues, such as studying the action presented here for Dirichlet
boundary conditions. Concrete calculations that could shed light
on this matter could be done concerning the issue of the conformal
anomaly of the boundary CFT, in the spirit of
refs.\cite{skenderis,deharo,skenderis1,papadimitriou}. One could
consider the situation in which $\overline{A}$ for AdS gravity
corresponds to the configuration used in our kounterterms
approach, and read the conformal anomaly from the variation of the
action under radial diffeos (which induce Weyl transformations in
the boundary for the boundary conditions of ref.\cite{motz1})with
$\overline{A}$ kept fixed. This would correspond to the picture of
the anomaly as arising from the non invariance of the regulator
(or subtraction procedure), as $\overline{A}$ in the our
kounterterms approach could be seen as a regulator.

One direction that seems both interesting and accessible has to do
with the extension to the supersymmetric case. It seems reasonable
to guess that the boundary terms that would result from the
transgression forms for suitable supersymmetric extensions of the
AdS groups would regulate, with proper coefficients relative to the
bulk, different versions of standard supergravities which are
extensions of AdS gravities. In this respect it is worthwhile to
remark that the issue of picking the right boundary terms in
standard supergravities related to the M-theory is an important one,
even with possible implications for phenomenology, as it can be seen
in the recent papers by Moss \cite{moss} and references therein. The
procedure followed in ref.\cite{moss} was to compute the boundary
terms order by order in an expansion in powers of some parameter,
which is not guaranteed to end.

It is also possible and worth exploring that the transgression
action of eq.(\ref{transgresion}) would be even better suited for
eventual application to the AdS-CFT correspondence, as its
invariance under gauge transformations for the AdS/conformal group
is exact (and of course off-shell) without further requirements or
conditions. That would not be surprising If one believes that the
effective field theory description of the M-theory with
corrections of higher order in the curvature included could in
fact be a Chern-Simons supergravity, which was originally proposed
in ref. \cite{troncoso2} and also explored in
refs.\cite{horava,banados2,nastase}.

Finally, the action has an obvious symmetry under the interchange
of $A$ and $\bar{A}$ and change of sign of the action, which may
have non trivial consequences worth exploring. It is intriguing
that Linde \cite{linde1,linde2} studied a model of gravity coupled
with scalar fields where the field content (including gravity)is
duplicated (as it is for us) and a similar symmetry, which Linde
calls {\it antipodal symmetry}, is the key to a way to solve the
cosmological constant problem\footnote{In ref.\cite{linde1,linde2}
both fields actually 'see each other' in the bulk indirectly
through the common volume element in the action, while in the
models discussed here they only see each other in the boundary.
However in the presence of branes with boundaries both fields
interact at the branes boundaries \cite{mn,mora}, so these could
mediate a bulk to bulk interaction. It is worthwhile to remark
that while 'antipodal symmetry' in Linde's model is postulated ad
hoc in the models discussed here it is a natural byproduct of the
construction of an action with enhanced symmetry.}.\\

{\bf Acknowledgments}\\ I am very grateful for discussions with R.
Olea, R. Troncoso and J. Zanelli on the subject of this paper, as
well as for the warm hospitality of the members of the Centro de
Estudios Cient\'{\i}ficos, C.E.C.S., of Valdivia, Chile during
several visits while this work was being done. I acknowledge funding
for my research from the Program 'Fondo Nacional de Investigadores'
(FNI, DINACYT, Uruguay).

\end{document}